\documentclass[twoside,11pt]{article}

\usepackage[english]{babel}

\usepackage{amsmath}
\usepackage{mathtools}

\usepackage{blindtext}
\usepackage{algpseudocode}
\usepackage{bm}
\usepackage{algorithm}
\usepackage{color}

\allowdisplaybreaks

\usepackage{placeins}
\usepackage{float}
\FloatBarrier

\newcommand{\numberthis}{\addtocounter{equation}{1}\tag{\theequation}}

\usepackage{jmlr2e}
\usepackage{hyperref}

\usepackage{lastpage}

\ShortHeadings{Storage capacity of perceptron with variable slection}{Xu, Ohzeki, and Kabashima}
\firstpageno{1}

\begin{document}

\title{Storage capacity of perceptron with variable selection}

\author{\name Yingying Xu\thanks{Corresponding author}\email 
yingying.xu@helsinki.fi \\
       \addr 
Department of Mathematics and Statistics, University of Helsinki, P.O. Box 68, FI-00014 Helsinki, Finland \\
Finnish Center for Artificial Intelligence(FCAI), Finland \\
RIKEN Center for Interdisciplinary Theoretical and Mathematical Sciences(iTHEMS), Wako, Saitama 351-0198, Japan
       \AND
       \name Masayuki Ohzeki \email masayuki.ohzeki.a4@tohoku.ac.jp \\
       \addr 
       Graduate School of Information Sciences, Tohoku University 
       Sendai 980-8579, Japan \\
       Department of Physics, 
       Institute of Science Tokyo 
       Tokyo 152-8551, Japan \\
       Sigma-i Co., Ltd. 
       Tokyo 108-0075, Japan 
       \AND
       \name Yoshiyuki Kabashima \email
       kaba@phys.s.u-tokyo.ac.jp \\
       \addr
       Institute for Physics of Intelligence,  
       The University of Tokyo, 7-3-1 Hongo, Bunkyo-ku, Tokyo 113-0033, Japan \\
       Department of Physics, 
       The University of Tokyo, 
       7-3-1 Hongo, Bunkyo-ku, Tokyo 113-0033, Japan \\
       Trans-Scale Quantum Science Institute, The University of Tokyo, 
       7-3-1 Hongo, Bunkyo-ku, Tokyo 113-0033, Japan
       }

\editor{}

\maketitle

\begin{abstract}
A central challenge in machine learning is to distinguish genuine structure from chance correlations in high-dimensional data. In this work, we address this issue for the perceptron, a foundational model of neural computation. Specifically, we investigate the relationship between the pattern load $\alpha$ and the variable selection ratio $\rho$ for which a simple perceptron can perfectly classify $P = \alpha N$ random patterns by optimally selecting $M = \rho N$ variables out of $N$ variables. While the Cover--Gardner theory establishes that a random subset of $\rho N$ dimensions can separate $\alpha N$ random patterns if and only if $\alpha < 2\rho$, we demonstrate that optimal variable selection can surpass this bound by developing a method, based on the replica method from statistical mechanics, for enumerating the combinations of variables that enable perfect pattern classification. This not only provides a quantitative criterion for distinguishing true structure in the data from spurious regularities, but also yields the storage capacity of associative memory models with sparse asymmetric couplings.
\end{abstract}


\section{Introduction}
In data analysis, determining whether a perfect classification genuinely reflects underlying structure or merely results from the model’s flexibility is a central statistical question.
From a hypothesis-testing perspective, the existence of a separating surface that achieves zero training error does not, by itself, constitute evidence of meaningful correlations.
One must ask whether the observed separability exceeds what could occur purely by chance.

This idea was first formalized by \citet{cover1965geometrical}, who computed the probability that $P$ randomly labeled points in $N$-dimensional space are linearly separable.
He showed that separability remains likely as long as $P < 2N$, defining a \textit{critical capacity} of roughly two patterns per degree of freedom.
Below this threshold, even random data can be perfectly separated, so zero error cannot be interpreted as evidence of structure; above it, perfect separability becomes exponentially unlikely for random data.
Cover’s result thus established a statistical boundary between chance fitting and genuine learning.

The complexity of feature interactions was later explored by \citet{Cover1977} in the context of the \textit{measurement-selection problem}.
They proved that, even under simple Gaussian assumptions, any monotone ordering of classification errors among feature subsets can in principle occur.
This means that the discriminative power of individual features does not predict that of their combinations—a weak pair of features may outperform a strong single one—and that no sequential selection rule is guaranteed to find the optimal subset.
From a statistical viewpoint, the mapping from marginal information to joint discriminability is therefore intrinsically non-monotonic.
An empirical counterpart of this theoretical anomaly was later reported by \citet{KenjiNagata2015}, who exhaustively analyzed neural data using sparse estimators such as 
{\em least absolute shrinkage and selection operator}
(LASSO) \citep{tibshirani1996regression} and 
{\em automatic relevance determination} (ARD)~\citep{MacKay1994ARD}.
They found that when the underlying data lack intrinsic discriminative information, feature selections become highly unstable, providing a finite-sample manifestation of Cover’s \textit{anomalous ordering}.
These works collectively emphasize that apparent separability or sparsity does not necessarily imply the presence of true information—it may merely reflect model flexibility within limited data.

A powerful framework for addressing these questions was developed by \citet{E.Gardner_1988} in the statistical-mechanical theory of the \textit{optimal storage capacity} of neural networks. 
By evaluating the typical volume of coupling space satisfying stability constraints for random patterns, they showed that 
the maximal number of patterns $P_{\rm c}$ that can be stored 
in associative memory models~\citep{Nakano1972,Amari1972,Kohonen1972,Hopfield1982} is determied by 
the Cover's capacity as $P_{\rm c}/N=2$. 
Their analysis demonstrated that storage capacity is determined by the entropy of feasible couplings—the volume of weight configurations compatible with all stored patterns under given constraints.
From this viewpoint, variable selection introduces a new type of constraint on the coupling space. Activating only a fraction $\rho$ of available input dimensions effectively restricts the network to a lower-dimensional manifold, analogous to imposing sparse connectivity or limited synaptic resources in associative memory models.
Thus, the problem of evaluating capacity under variable selection is closely related to determining the storage capacity of \textit{constrained associative memory models}, where patterns must be stored using a restricted subset of synapses.
In both cases, the critical capacity reflects how the entropy of feasible configurations changes under structural constraints.

Motivated by these insights, the present study develops a \textit{statistical-mechanical theory of perceptrons with variable selection}.
By analyzing the typical volume of weight configurations that correctly classify random patterns while activating only a fraction $\rho$ of input dimensions, we quantify how such structural restrictions reshape 
the classical Cover--Gardner theory.
Our formulation unifies three perspectives—Cover’s geometrical separability, Gardner’s optimal-storage theory, and the modern view of sparse associative memory—within a single framework for understanding how structural constraints control the boundary between random separability and meaningful representation.

The present paper is organized as follows. 
In the next section, we formulate the problem that this paper aims to address.
Section 3 outlines the proposed method, and Section 4 presents the explicit computational procedure based on the replica method.
Section 5 reports the analytical results, and Section 6 verifies these results through numerical experiments.
Section 7 is devoted to the conclusion and discussion.

\section{Problem setting}

As a general setting, let us suppose a situation where for each of $P$ input vectors $\bm{x}_\mu $  of $N$-dimension ($\mu=1,\ldots, P$), which are assumed to be sampled independently and uniformly from $\{+1,-1\}^N$ or the $N$-dimensional sphere centered at the origin, binary label $y_\mu$ is assigned independently and uniformly from $\{+1,-1\}$.
Our goal is to evaluate the maximal value of \(P\), $P_{\rm c}$, for which there exists 
a simple perceptron with weight vector \(\bm{w}\in\mathbb{R}^{N}\) that correctly 
reproduces all labels, that is,
\begin{align*}
y_{\mu}
=
\operatorname{sign}\!\left(
\frac{1}{\sqrt{N}}
\sum_{i=1}^{N} w_{i} x_{\mu i}
\right),
\qquad \mu = 1,\ldots,P,
\numberthis \label{eq:separable}
\end{align*}
under the sparsity constraint that the number of nonzero components of 
\(\bm{w}\) is \(N\rho\), where \(0 < \rho \le 1\), 
for typical random datasets $\xi^P$.

When the nonzero components of the weight vector are chosen at random,
the Cover--Gardner theory immediately gives \(P_{\rm c}/N = 2\rho\).
Our question, however, is fundamentally different: 
\emph{how large can \(P_{\rm c}\) become when the optimal combination of nonzero components is selected?}
This type of question naturally arises when classifying data into two classes
using \(N\) experimentally obtained features.
In many experiments, one does not know in advance which of the \(N\) features
are actually relevant for the classification task.
Thus, one typically searches for a subset of features that yields the most 
``regular-looking'' separation of the two classes.
However, it then becomes crucial to determine whether the identified regularity
genuinely reflects an underlying structure in the data, or whether it merely
appears regular due to accidental patterns that can emerge from random labeling.
Our question provides a quantitative criterion for distinguishing true structure 
in the data from spurious regularity.

This problem can also be interpreted as evaluating the performance of a sparsely 
connected associative memory model.
Let \(\bm{x} \in \{+1,-1\}^{N}\) represent the states of \(N\) binary neurons and
\(y \in \{+1,-1\}\) represent the state of an \(N+1\)-th binary neuron, with 
\(w_i\) interpreted as the synaptic connection linking neuron \(i\in \{1,\ldots, N\} \) to neuron \(N+1\).
Under this correspondence, condition~(\ref{eq:separable}) expresses the requirement
that an associative memory model composed of \(N+1\) binary neurons, each having only 
\(N\rho\) synaptic connections, can store \(P\) random 
patterns as stable memory states.
Thus, \(P_{\rm c}\) represents the storage capacity of such an associative memory model 
with sparse, asymmetric synaptic connectivity.

\begin{figure}
\centering
\includegraphics[keepaspectratio,width=14cm]{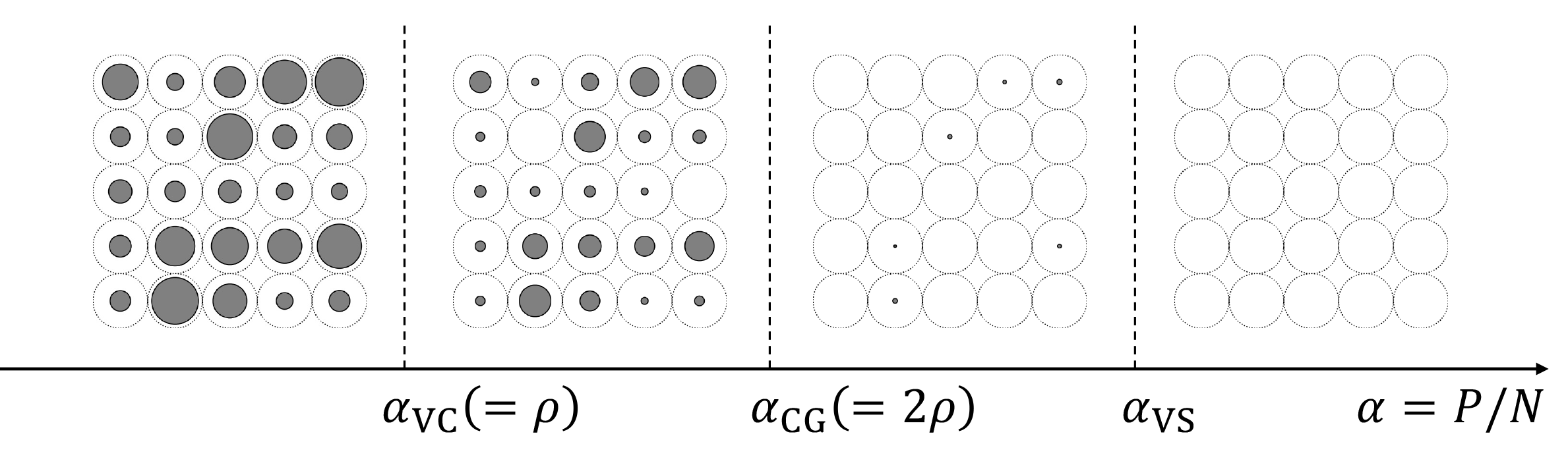}
\caption{\label{fig:phase_graph_capacity}
Schematic illustration of how the capacity is determined for a fixed variable selection ratio $\rho$.  
The vector $\bm{c}$ specifies a cluster defined by a particular choice of selected variables, 
represented by a circle of dotted line.  
Shaded regions indicate the feasible regions compatible with $\xi^{P}$.
For $\alpha < \alpha_{\rm VC} = \rho$, corresponding to the 
Vapnik--Chervonenkis bound \citep{Vapnik1971}, all clusters possess feasible regions of finite volums for typical random datasets $\xi^P$.  
For $\alpha_{\rm VC} < \alpha < \alpha_{\rm CG} = 2\rho$, 
although a small fraction of clusters disappears, 
typical clusters still retain feasible regions of finite volume.  
For $\alpha_{\rm CG} < \alpha < \alpha_{\rm VS}$, 
typical clusters vanish, yet an exponential number of atypical clusters continue to have nonzero feasible volumes.  
For $\alpha > \alpha_{\rm VS}$, the feasible region disappears in all clusters. 
The goal of the present work is to evaluate $\alpha_{\rm VS}$. 
}
\end{figure}


\section{Analytical formulation}
To explicitly represent the sparsity constraint of the simple perceptron,
we introduce a binary vector $\bm{c} = (c_i) \in \{0,1\}^N$ and rewrite
Eq.~(\ref{eq:separable}) as
\begin{align*}
y_{\mu}
=
\operatorname{sign}\!\left(
\frac{1}{\sqrt{N}}
\sum_{i=1}^{N} c_i w_i x_{\mu i}
\right),
\qquad
\mu = 1,\ldots,P .
\numberthis \label{eq:sparse_separable}
\end{align*}
For a fixed choice of nonzero components $\bm{c}$, we then evaluate the
volume of weight vectors $\bm{w}$ compatible with
Eq.~(\ref{eq:sparse_separable}) under the norm constraint
$\sum_{i=1}^{N} c_i w_i^{2} = N\rho$.
For this calculation, it is convenient to introduce the improper
conditional distribution \citep{Kuhlmann1994}
\begin{align*}
P(\bm{w}\mid \bm{c})
=
\frac{1}{(2\pi)^{N/2}}
\exp\!\left(
        -\sum_{i=1}^{N} \frac{1-c_i}{2}\, w_i^{2}
    \right)
\end{align*}
which yields the volume
\begin{align*}
V(\bm{c} \mid \xi^{P})
=
\int d\bm{w}\,
P(\bm{w}\mid \bm{c})
\prod_{\mu=1}^{P}
\Theta\!\left(
    \frac{y_{\mu}}{\sqrt{N}}
    \sum_{i=1}^{N} c_i w_i x_{\mu i}
\right)
\delta\!\left(
    \sum_{i=1}^{N} c_i w_i^{2} - N\rho
\right),
\numberthis \label{eq:Vc}
\end{align*}
where
\(
\xi^{P}
=
\{(\bm{x}_{1},y_{1}),\ldots,(\bm{x}_{P},y_{P})\}
\)
and 
\begin{align*}
    \Theta(x) = \left \{
    \begin{array}{ll}
    1, & x\ge 0, \cr
    0, & \mbox{otherwise}.
    \end{array}
    \right .
\end{align*}

For typical datasets $\xi^{P}$, we aim to determine the largest value
of \(P\), $P_{\rm c}$, for which one can choose a vector \(\bm{c}\) satisfying
\(V(\bm{c}\mid \xi^{P}) > 0\).
For this purpose, we use the identity
\begin{align*}
\lim_{m\to 0} V^{m}(\bm{c}\mid \xi^{P})
=
\begin{cases}
1, & V(\bm{c}\mid \xi^{P}) > 0, \\[1mm]
0, & V(\bm{c}\mid \xi^{P}) = 0,
\end{cases}
\end{align*}
which implies that the total number of $\bm{c}$ satisfying
$V(\bm{c}\mid \xi^{P}) > 0$ can be written as
\begin{align*}
\mathcal{N}(\xi^{P})
=
\lim_{m\to 0}
\sum_{\bm{c}}
V^{m}(\bm{c}\mid \xi^{P})
\,
\delta\!\left(
    \sum_{i=1}^{N} c_i - N\rho
\right).
\end{align*}

The quantity $\mathcal{N}(\xi^{P})$ fluctuates depending on the
realization of $\xi^{P}$.
However, according to the large deviation property of the number of combinations, it is reasonable to assume the scaling form
\(
\mathcal{N}(\xi^{P}) \sim \exp(N s)
\)
with probability
\(
P(s) \simeq \exp[-N I(s)]
\)
\citep{Monasson1994,Engel1996}. 
The typical value of \(s\), which minimizes \(I(s)\) to zero, is
then given by
\begin{align*}
\Sigma&=\mathbb{E}_{\xi^{P}}[s]
=
\frac{1}{N}
\mathbb{E}_{\xi^{P}}[\ln \mathcal{N}(\xi^{P})]
=
\lim_{n\to 0}
\frac{\partial}{\partial n}
\frac{1}{N}
\ln \mathbb{E}_{\xi^{P}}[\mathcal{N}^{\,n}(\xi^{P})]
\cr
&=
\lim_{m\to 0}\,
\lim_{n\to 0}
\frac{\partial}{\partial n}
\frac{1}{N}
\ln
\mathbb{E}_{\xi^{P}}
\left[
    \left(
        \sum_{\bm{c}}
        V^{m}(\bm{c}\mid \xi^{P})
        \,
        \delta\!\left(
            \sum_{i=1}^{N} c_i - N\rho
        \right)
    \right)^{n}
\right], 
\numberthis \label{eq:typical_entropy}
\end{align*}
where $\mathbb{E}_X[ \cdots ]$ generally stands for 
the average operation with respect to $X$. 

 The entropy density $s=N^{-1}\ln{\mathcal N}(\xi^P)$ cannot be negative because ${\mathcal N}(\xi^P)$ is a natural number.
Therefore, the critical value 
$P_{\rm c}$ can be determined as the point where
its typical value $\Sigma$ vanishes (Fig.~\ref{fig:phase_graph_capacity}).

\section{Replica computation}
\label{replica_calculations}
Unfortunately, it is difficult to evaluate 
Eq.~(\ref{eq:typical_entropy}) in a mathematically rigorous manner.
To circumvent this difficulty in practice, we employ the 
non-rigorous replica method from statistical mechanics 
\citep{Mezard1987}.
This method consists of the following two steps:
\begin{enumerate}
  \item For positive integers \(n\) and \(m\), evaluate
  \begin{align*}
    \phi(n,m)
    &= \frac{1}{N}
       \ln 
       \mathbb{E}_{\xi^{P}}
       \biggl[
         \biggl(
           \sum_{\bm{c}}
           V^{m}(\bm{c}\mid \xi^{P})
           \,
           \delta\!\left(
             \sum_{i=1}^{N} c_i - N\rho
           \right)
         \biggr)^{n}
       \biggr] ,
  \end{align*}
  as a function of \(n\) and \(m\).

  \item Compute
  \begin{align*}
    \numberthis \label{eq:Sigma_replica}
    \Sigma
    &= \lim_{m\to 0}\,
       \lim_{n\to 0}
       \frac{\partial \phi(n,m)}{\partial n},
  \end{align*}
  by analytically continuing the resulting expression 
  of \(\phi(n,m)\) to real values \(n,m \in \mathbb{R}\).
\end{enumerate}
Details of these steps are provided below. 

\subsection{Computation of \texorpdfstring{$\phi(n,m)$}{phi(n,m)} for integers $n$ and $m$}

Substituting Eq.~(\ref{eq:Vc}) into Eq.~(\ref{eq:typical_entropy}) yields
\begin{align*}
  &\mathbb{E}_{\xi^{P}}
  \biggl[
    \biggl(
      \sum_{\bm{c}}
      V^{m}(\bm{c}\mid \xi^{P})
      \,
      \delta\!\left(
        \sum_{i=1}^{N} c_i - N\rho
      \right)
    \biggr)^{n}
  \biggr]
  \notag\\
  &\quad
  = \sum_{\bm{c}^1,\ldots,\bm{c}^n }
    \int \prod_{a=1}^{n}\prod_{\sigma=1}^{m} d\bm{w}^{a\sigma}
    \biggl\{
      \underbrace{
        \mathbb{E}_{\xi^P}
        \biggl[
          \prod_{a=1}^{n}\prod_{\sigma=1}^{m}
          \prod_{\mu = 1}^P
          \Theta \!\left(
            \frac{y_\mu}{\sqrt{N}} 
            \sum_{i=1}^N c_i^{a}w_i^{a\sigma} x_{\mu i}
          \right)
        \biggr]
      }_{A}  
      \notag\\
  &\qquad\qquad\times 
      \underbrace{
        \prod_{a=1}^{n}\prod_{\sigma = 1}^{m} 
        \biggl(
          P(\bm{w}^{a\sigma}\mid \bm{c}^a)\,
          \delta \!\left(
            \sum_{i=1}^N c_i^a (w_i^{a\sigma})^2 -N \rho
          \right)
        \biggr)
        \prod_{a=1}^n 
        \delta \!\left(
          \sum_{i=1}^N c_i^a -N\rho
        \right)
      }_{B}
    \biggr\}.
     \numberthis \label{eq:V_finite_n_m}
\end{align*}
We evaluate the contributions of \(A\) and \(B\) separately. 

\paragraph{Contribution \(A\).}
The quantity \(A\) is evaluated using the following facts:
\begin{itemize}
  \item
  The input vectors \(\bm{x}_1,\ldots,\bm{x}_{P}\)
  are independently drawn from the uniform distribution over
  \(\{+1,-1\}^N\) or from the \(N\)-dimensional sphere.  
  For each \(\bm{x}_{\mu}\), the label \(y_{\mu}\) is also drawn from  
  \(\{+1,-1\}\) uniformly.
  Thus, \(A\) is obtained by averaging
  \[
    \Theta \!\left(
      \frac{y}{\sqrt{N}} 
      \sum_{i=1}^N c_i^{a}w_i^{a\sigma} x_i
    \right)
  \]
  with respect to a single pair \((\bm{x},y)\), and raising the result
  to the \(P\)-th power.

  \item
  For \(\bm{x}\) uniformly distributed over
  \(\{+1,-1\}^N\) or the \(N\)-dimensional sphere,  
  the central limit theorem implies that
  \[
    u^{a\sigma} 
    = \frac{y}{\sqrt{N}} \sum_{i=1}^N c_i^{a}w_i^{a\sigma} x_i
    \quad
    (a=1,\ldots,n,\; \sigma=1,\ldots,m)
  \]
  follow a zero-mean multivariate normal
  distribution with covariance
  \begin{align*}
    \mathbb{E}_{\bm{x},y}[u^{a\sigma} u^{b\tau}]
    &= \frac{1}{N} 
       \sum_{i=1}^N 
       (c_i^a w_i^{a\sigma})
       (c_i^b w_i^{b\tau})
       \notag\\
    &=: q_{ab;\sigma \tau},
      \numberthis \label{eq:covariances}
  \end{align*}
  independently of \(y\).
\end{itemize}

Using these observations, we obtain
\begin{align*}
  A 
  &= \left(
    \int 
    \frac{
      \displaystyle 
      \prod_{a=1}^{n}\prod_{\sigma = 1}^{m}du^{a\sigma}
      \exp\!\left(
        -\frac{1}{2}\bm{u}^\top \mathcal{Q}^{-1} \bm{u}
      \right)
    }{
      (2\pi)^{nm/2} 
      \bigl(\det \mathcal{Q}\bigr)^{1/2}
    }
    \prod_{a=1}^{n}\prod_{\sigma =1}^{m}
      \Theta(u^{a\sigma})
  \right)^P,
\end{align*}
where \(\bm{u}=(u^{a\sigma})\) and 
\(\mathcal{Q}\) denotes the \(nm \times nm\) matrix composed of 
\(q_{ab;\sigma \tau}\). 

\paragraph{Contribution \(B\).} The contribution \(B\) is handled together with the
volume of the subshell of configurations
\(\bm{c}^1,\ldots,\bm{c}^n\), 
\(\bm{w}^{11}, \ldots, \bm{w}^{nm}\) that satisfy fixed order parameters 
\(q_{ab;\sigma\tau}\)
\((a,b \in \{1,\ldots,n\},\; \sigma, \tau \in \{1,\ldots,m\})\).
Specifically, we insert the identities
\begin{align*}
  1 
  &= N\int_{-\infty}^{+\infty} dq_{ab;\sigma\tau} 
     \,\delta
     \biggl(
       \sum_{i=1}^N c_i^a c_i^b w_i^{a\sigma}w_i^{b\tau}
       -Nq_{ab;\sigma\tau}
     \biggr)
     \notag\\
  &= \frac{N}{2\pi}
     \int_{-\infty}^{+\infty} 
       dq_{ab;\sigma\tau} 
     \int_{-i\infty}^{+i\infty}
       d\hat{q}_{ab;\sigma\tau}
     \exp \biggl[
       \hat{q}_{ab;\sigma\tau}
       \biggl(
         \sum_{i=1}^N c_i^a c_i^b w_i^{a\sigma}w_i^{b\tau}
         -Nq_{ab;\sigma\tau}
       \biggr)
     \biggr],
\end{align*}
\begin{align*}
  \delta \!\left(
    \sum_{i=1}^N c_i^a (w_i^{a \sigma})^2 -N \rho
  \right)
  &= \frac{1}{4\pi} 
     \int_{-i\infty}^{+i\infty}
       d\hat{q}_{aa;\sigma \sigma}
     \exp\biggl[
       -\frac{\hat{q}_{aa;\sigma \sigma}}{2}
       \biggl(
         \sum_{i=1}^N (c_i^a w_i^{a\sigma})^2 -N\rho 
       \biggr)
     \biggr],
\end{align*}
and 
\begin{align*}
  \delta \!\left(
    \sum_{i=1}^N c_i^a -N\rho
  \right)
  &= \frac{1}{2\pi} 
     \int_{-i\infty}^{+i\infty} dK_a
     \exp\biggl[
       -K_a 
       \biggl(
         \sum_{i=1}^N c_i^a -N\rho
       \biggr)
     \biggr],
\end{align*}
into Eq.~(\ref{eq:V_finite_n_m}), and perform the summation and integration 
over all possible configurations
\(\bm{c}^1,\ldots, \bm{c}^n\) and 
\(\bm{w}^{11}, \ldots, \bm{w}^{nm}\). 
This yields
\begin{align*}
  &\sum_{\bm{c}^1,\ldots,\bm{c}^n}
    \int \prod_{a=1}^{n}\prod_{\sigma =1 }^{m}
    d\bm{w}^{a\sigma}
    \exp\biggl[
      -\sum_{a=1}^{n}K_a \sum_{i=1}^N c_i^a 
      + 
      \sum_{a \le b}\sum_{\sigma \le \tau} 
      \hat{q}_{ab;\sigma \tau }
      \sum_{i=1}^N c_i^a c_i^b w_i^{a\sigma}w_i^{b\tau}
    \biggr]
    \times B
    \notag\\
  &\quad=
    \left(
      \frac{1}{(2\pi)^{nm/2}}
      \sum_{c^1,\ldots,c^n}
      \int 
      \prod_{a=1}^{n}\prod_{\sigma = 1}^{m} dw^{a\sigma}
      \exp\biggl[
        -\sum_{a=1}^n K_a c^a 
        +\mathcal{L}
           \bigl(\{c^a\}, \{w^{a\sigma}\}, 
                 \{\hat{q}_{ab;\sigma\tau}\}\bigr)
      \biggr]
    \right)^N, 
\end{align*}
where 
\begin{align*}
  &\mathcal{L}
    \bigl(\{c^a\}, \{w^{a\sigma}\}, 
          \{\hat{q}_{ab;\sigma\tau}\}\bigr)
    \notag\\
  &\quad
  = -\sum_{a=1}^n \frac{1-c^a}{2}
      \sum_{\sigma=1}^m (w^{a\sigma})^2
    -\sum_{a=1}^{n}\sum_{\sigma=1}^{m} 
      \frac{\hat{q}_{aa;\sigma\sigma} (c^a w^{a\sigma})^2}{2}
    + \sum_{\substack{a \le b, \sigma \le \tau \\[1pt]
                      a\ne b \,\vee\, \sigma \ne \tau }}
      \hat{q}_{ab;\sigma\tau} c^a c^b w^{a\sigma} w^{b\tau}.
\end{align*}

For \(N\gg 1\), substituting these into 
Eq.~(\ref{eq:V_finite_n_m}) and employing 
the saddle-point method 
provides an expression of \(\phi(n,m)\) as 
\begin{align*}
  \phi(n,m) 
  &=
  \mathop{\mathrm{extr}}_{
    \{K^a\}, \{q_{ab;\sigma\tau}\}
    ,\{\hat{q}_{ab;\sigma\tau}\}
  }
  \biggl\{
    \alpha \ln\biggl[
      \int 
      \frac{
        \prod_{a=1}^{n}\prod_{\sigma = 1}^{m}du^{a\sigma}\,
        \exp\!\left(
          -\frac{1}{2}\bm{u}^\top \mathcal{Q}^{-1}\bm{u}
        \right)
      }{
        (2\pi)^{nm/2} \bigl(\det \mathcal{Q} \bigr)^{1/2}
      }
      \prod_{a=1}^{n}\prod_{\sigma =1}^{m}
        \Theta(u^{a\sigma})
    \biggr]
    \notag\\
  &
    + \ln\biggl[
      \frac{1}{(2\pi)^{nm/2}} \!\!
      \sum_{c^1,\ldots,c^n} \!
      \int 
      \prod_{a=1}^{n}\prod_{\sigma = 1}^{m} dw^{a\sigma}
      \exp\biggl(
        -\sum_{a=1}^n K_a c^a 
        +\mathcal{L}
           \bigl(\{c^a\}, \{w^{a\sigma}\}, 
                 \{\hat{q}_{ab;\sigma\tau}\}\bigr)
      \biggr)
    \biggr]
    \notag\\
  &
    + \rho\sum_{a=1}^n K_a 
    + \rho \sum_{a=1}^{n}\sum_{\sigma=1}^{m}
      \frac{\hat{q}_{aa;\sigma\sigma}}{2}
    -\sum_{\substack{a \le b, \sigma \le \tau \\[1pt]
                     a\ne b \,\vee\, \sigma \ne \tau }}
      \hat{q}_{ab;\sigma\tau} q_{ab;\sigma\tau} 
  \biggr\},
  \numberthis \label{eq:logV_integers_n_m}
\end{align*}
for integers \(n\) and \(m\), where 
\(\alpha = P/N\) and 
\(\mathop{\mathrm{extr}}_X\{ f(X) \}\) denotes 
extremization of \(f(X)\) with respect to \(X\).

\subsection{Replica symmetry and analytical continuation to \texorpdfstring{$n,m \in \mathbb{R}$}{n,m in R}}

Next, we analytically continue Eq.~(\ref{eq:logV_integers_n_m}) to 
real values \(n,m \in \mathbb{R}\). 
Replica symmetry, i.e., the invariance of the right-hand side of 
Eq.~(\ref{eq:V_finite_n_m}) under any permutation of the replica indices 
\(a \in \{1,\ldots,n\}\) and \(\sigma \in \{1,\ldots,m\}\),
plays a key role in this operation. 
Since the exact computation of Eq.~(\ref{eq:V_finite_n_m}) possesses this property, it is natural to assume that 
the extremum of the right-hand side of 
Eq.~(\ref{eq:logV_integers_n_m}) also exhibits
the same symmetry. 
Therefore, we perform the extremization
assuming that the order parameters are of the form
\begin{align*}
\numberthis \label{eq:RS_assumption}
  q_{ab;\sigma \tau} 
  &= 
  \begin{cases}
    \rho, & a=b,\ \sigma = \tau, \\
    q_1,  & a=b,\ \sigma \ne \tau, \\
    q_0,  & a\ne b,
  \end{cases}
  \qquad
  \hat{q}_{ab;\sigma \tau} 
  = 
  \begin{cases}
    \hat{Q},   & a=b,\ \sigma = \tau, \\
    \hat{q}_1, & a=b,\ \sigma \ne \tau, \\
    \hat{q}_0, & a\ne b,
  \end{cases}
  \qquad
  K^a = K .
\end{align*}

Under this assumption, we obtain
\begin{align*}
  \phi(n,m)
  &= \mathop{\mathrm{extr}}_{
        q_1,q_0, 
        \hat{Q},\hat{q}_1, \hat{q}_0, K
      }
    \biggl\{
      \alpha \ln
      \biggl[
        \int Dz
        \left(
          \int Dy 
          H\!\left(
            \frac{
              \sqrt{q_1-q_0}\,y + \sqrt{q_0}\, z
            }{
              \sqrt{\rho-q_1}
            }
          \right)^m 
        \right)^n
      \biggr]
      \notag\\
  &\qquad
    + \ln\biggl[
      \int Dz 
      \left(
        1 + 
        \frac{e^{-K}}{ \bigl(\hat{Q}+\hat{q}_1\bigr)^{m/2}} 
        \int Dy 
        \exp\!\left(
          \frac{
            m\bigl(\sqrt{\hat{q}_1-\hat{q}_0}\,y
                   +\sqrt{\hat{q}_0}\,z \bigr)^2
          }{
            2\bigl(\hat{Q}+\hat{q}_1 \bigr)
          }
        \right) 
      \right)^n
    \biggr]
      \notag\\
  &\qquad
    + nK\rho 
    + \frac{nm}{2}\hat{Q}\rho 
    -\frac{nm(m-1)}{2}
      \bigl(\hat{q}_1q_1 -\hat{q}_0 q_0 \bigr)
    -\frac{nm(nm-1)}{2}\hat{q}_0 q_0
    \biggr\},
    \numberthis \label{eq:RS_generating_function}
\end{align*}
where \(Dz = dz \exp(-z^2/2)/\sqrt{2\pi}\) 
generally denotes the standard Gaussian measure and 
\(H(x) = \int_x^{+\infty} Dz\).
Its derivation is shown in Appendix~\ref{app:RS_computation}. 
This expression is well-defined for 
\(n,m \in \mathbb{R}\). 
Therefore,
we can evaluate Eq.~(\ref{eq:Sigma_replica}) 
using Eq.~(\ref{eq:RS_generating_function}), 
which yields
\begin{align*}
    \phi(m) 
    &= \lim_{n\to 0}\frac{\partial \phi(n,m)}{\partial n} 
    \notag\\
    &=\mathop{\mathrm{extr}}_{
        q_1,q_0, 
        \hat{Q},\hat{q}_1, \hat{q}_0, K
      }
    \biggl\{
      \alpha \int Dz\,
      \ln
        \left(
          \int Dy 
          H\!\left(
            \frac{
              \sqrt{q_1-q_0}\,y + \sqrt{q_0}\, z
            }{
              \sqrt{\rho-q_1}
            }
          \right)^m 
        \right)
      \notag\\
  &\qquad
    + \int Dz \,\ln
      \left(
        1 + 
        \frac{e^{-K}}{ \bigl(\hat{Q}+\hat{q}_1\bigr)^{m/2}} 
        \int Dy 
        \exp\!\left(
          \frac{
            m\bigl(\sqrt{\hat{q}_1-\hat{q}_0}\,y
                   +\sqrt{\hat{q}_0}\,z \bigr)^2
          }{
            2\bigl(\hat{Q}+\hat{q}_1 \bigr)
          }
        \right) 
      \right)
      \notag\\
  &\qquad
    + K\rho 
    + \frac{m}{2}\bigl(\hat{Q}\rho+\hat{q}_1q_1\bigr) 
    -\frac{m^2}{2}
      \bigl(\hat{q}_1q_1 -\hat{q}_0 q_0 \bigr)
    \biggr\}, 
    \numberthis \label{eq:RS_free_energy_n_zero}
\end{align*}
and \(\Sigma = \lim_{m\to 0}\phi(m)\). 

\section{Results}

The extremization condition of Eq.~(\ref{eq:RS_free_energy_n_zero}) is given by
\begin{align*}
    \hat{q}_1 
    &= 
    \frac{\alpha}{\rho-q_1} \int Dz\,
    \frac{ 
      \displaystyle \int Dy\, H^m
      \left (\frac{H^\prime}{H} \right)^2 
    }{
      \displaystyle \int Dy\, H^m 
    }, 
    \numberthis \label{eq:RS_SP_qh1}\\[3pt]
    \hat{q}_0 
    &= 
    \frac{\alpha}{\rho-q_1}\int Dz\,
    \left(
      \frac{ 
        \displaystyle \int Dy\, H^m \frac{H^\prime}{H} 
      }{ 
        \displaystyle \int Dy\, H^m 
      }
    \right)^2, 
    \numberthis \label{eq:RS_SP_qh0} \\[3pt]
    \rho 
    &= \frac{\rho}{\hat{Q}+\hat{q}_1}
    + q_1, 
    \numberthis \label{eq:RS_SP_Q} \\[3pt]
    q_1 
    &= \int Dz\,
    \frac{e^{-K} \displaystyle \int Dy\, \Xi^m \omega^2 }
         {1+e^{-K}\displaystyle \int Dy\, \Xi^m},
     \numberthis \label{eq:RS_SP_q1}\\[3pt]
    q_0 
    &= \int Dz\,
    \left(
      \frac{e^{-K} \displaystyle \int Dy\, \Xi^m \omega}
           {1+e^{-K}\displaystyle \int Dy\, \Xi^m}
    \right)^2,
    \numberthis \label{eq:RS_SP_q0} \\[3pt]
    \rho 
    &= \int Dz\,
    \frac{e^{-K} \displaystyle \int Dy\, \Xi^m}
         {1+ e^{-K}\displaystyle \int Dy\, \Xi^m},
    \numberthis \label{eq:RS_SP_rho}
\end{align*}
where 
\[
\omega = 
\frac{\sqrt{\hat{q}_1-\hat{q}_0}\,y+\sqrt{\hat{q}_0}\,z}
     {\hat{Q}+\hat{q}_1},
\qquad
\Xi = 
\bigl(\hat{Q}+\hat{q}_1 \bigr)^{-1/2}
\exp\!\left(
  \frac{
    \bigl(\sqrt{\hat{q}_1-\hat{q}_0}\,y
           +\sqrt{\hat{q}_0}\,z \bigr)^2
  }{
    2\bigl(\hat{Q}+\hat{q}_1 \bigr)
  }
\right).
\]
The solution of these equations for \(m\to 0\) 
is classified into two cases depending on \(\alpha\).

\subsection{\texorpdfstring{$\alpha < \alpha_{\rm CG}(=2\rho)$}{alpha < alpha^typ (=2 rho)}}

When \(\alpha\) is sufficiently small, 
\(\hat{q}_1, \hat{q}_0 = O(1)\) and  
\(0<q_0< q_1<\rho\) hold.  
In this regime, taking \(m\to 0\) yields 
\(H^m\to 1\) and \(\Xi^m \to 1\), 
which reduces Eqs.~(\ref{eq:RS_SP_qh1})–(\ref{eq:RS_SP_rho}) to
\begin{align*}
    \hat{q}_1 
    &= \frac{\alpha}{\rho-q_1}
    \int Dz\, Dy\,
    \left(\frac{H^\prime
    \left (
    \frac{
    \sqrt{q_1-q_0}y + \sqrt{q_0}z
    }{\sqrt{\rho-q_1}}
    \right )
    }{H
    \left (
    \frac{
    \sqrt{q_1-q_0}y + \sqrt{q_0}z
    }{\sqrt{\rho-q_1}}
    \right )
    }\right)^2 
    =\frac{\alpha}{\rho-q_1}\int Dt 
    \left(
      \frac{H^\prime (\gamma t)}
           {H (\gamma t)}
    \right)^2,
    \numberthis \label{eq:RS_SP_qh1_2}\\[3pt]
    \hat{q}_0 
    &= \frac{\alpha}{\rho-q_1}
    \int Dz\,
    \left(
      \int Dy\, \frac{H^\prime
      \left (
    \frac{
    \sqrt{q_1-q_0}y + \sqrt{q_0}z
    }{\sqrt{\rho-q_1}}
    \right )
      }{H
      \left (
    \frac{
    \sqrt{q_1-q_0}y + \sqrt{q_0}z
    }{\sqrt{\rho-q_1}}
    \right )
      }
    \right)^2, 
    \numberthis \label{eq:RS_SP_qh0_2}\\[3pt]
    \rho 
    &= \frac{\rho}{\hat{Q}+\hat{q}_1} + q_1, 
    \numberthis \label{eq:RS_SP_Q_2}\\[3pt]
    q_1 
    &= \frac{e^{-K}}{1+e^{-K}}
    \int Dz\, Dy\, \omega^2 
    = \frac{\rho \hat{q}_1}{(\hat{Q}+\hat{q}_1)^2},
    \numberthis \label{eq:RS_SP_q1_2}\\[3pt]
    q_0 
    &= \int Dz 
    \left(
      \frac{e^{-K}}{1+e^{-K}} 
      \int Dy\, \omega
    \right)^2
    =\frac{\rho^2\hat{q}_0}{(\hat{Q}+\hat{q}_1)^2},
    \numberthis \label{eq:RS_SP_q0_2}\\[3pt]
    \rho 
    &= \frac{e^{-K}}{1+e^{-K}},
    \numberthis \label{eq:RS_SP_rho_2}
\end{align*}
where \(\gamma = \sqrt{q_1/(\rho-q_1)}\). 
In addition, applying \(m\to 0\) in 
Eq.~(\ref{eq:RS_free_energy_n_zero}) gives
\begin{align*}
    \Sigma = \lim_{m\to 0}\phi(m) 
    =-(1-\rho)\ln(1-\rho)-\rho\ln\rho,   
\end{align*}
which coincides with the entropy density 
for selecting \(N\rho\) variables 
out of \(N\) variables. 
This implies that for any choice of 
\(N\rho\) variables there exists a simple perceptron 
that is consistent with the given random patterns
$\xi^P=\{(\bm{x}_1,y_1),\ldots,(\bm{x}_P, y_P)\}$. 

The overlap \(q_1\) grows 
as \(\alpha\) increases. 
Eqs.~(\ref{eq:RS_SP_qh1_2}) and 
(\ref{eq:RS_SP_q1_2}), together with 
the asymptotic forms \(H^\prime(x)/H(x) \simeq 
-x\) for \(x\gg 1\) and \(0\) for \(x\ll -1\), 
indicate that 
\begin{align*}
    \frac{\rho q_1 }{(\rho-q_1)^2} 
    \simeq 
    \frac{\alpha q_1}{(\rho-q_1)^2}
    \int Dt \Theta(t) t^2 
    = \frac{\alpha q_1}{2(\rho-q_1)^2}
\end{align*}
holds when \(q_1\to \rho\) from below, 
which defines a critical pattern ratio
\begin{align*}
\alpha_{\rm CG} = 2\rho.    
\end{align*}

The Cover–Gardner theory guarantees that, in typical cases, simple perceptrons can correctly separate 
random patterns as long as the number of patterns does not exceed twice the dimension of the input vectors. The present analysis reproduces this well-known result within the new formulation that incorporates variable selection.

\subsection{$\alpha > \alpha_{\rm CG}$}
For $\alpha > \alpha_{\rm CG}$, $\hat{q}_1$ and $\hat{q}_0$ diverge, and 
$q_1$ converges to $\rho$ in $m\to 0$. Therefore, 
we rescale relevant variables as
\begin{align*}
    F_1 = m^2 \hat{q}_1, \quad F_0 = m^2\hat{q}_0, 
    \quad E=m(\hat{Q}+\hat{q}_1), 
    \quad \chi=\frac{\rho -q_1}{m}.
\end{align*}
Accordingly, we have 
\begin{align*}
   &\lim_{m\to 0} H^m\left (
    \frac{\sqrt{q_1-q_0}y+\sqrt{q_0}z}{
    \sqrt{\rho -q_1}} \right )
    = \Theta(-v)+ \Theta(v) e^{-v^2/(2\chi)}=:\tilde{H}(v, \chi), \cr
&\lim_{m\to 0}\Xi^m = \exp\left (\frac{h^2}{2E} \right ) =: \tilde{\Xi}(h, E), \notag
\end{align*}
in this limit, 
where $v = \sqrt{\rho-q_0}y + \sqrt{q_0}z$
and $h = \sqrt{F_1-F_0}y + \sqrt{F_0} z$. Then, 
Eqs.~(\ref{eq:RS_SP_qh1})--(\ref{eq:RS_SP_rho})
are rewritten as 
\begin{align}
    F_1 &= \frac{\alpha}{\chi^2}
    \int Dz
    \frac{\int Dy \tilde{H}(v, \chi) \Theta(v) v^2 }{\int Dy \tilde H(v,\chi)}, \\
    F_0 &=\frac{\alpha}{\chi^2}
    \int Dz \left (\frac{\int Dy \tilde{H}(v,\chi) \Theta(v) v}
    {\int Dy \tilde{H}(v,\chi)} \right )^2, \\
    \rho &= \frac{1}{E^2}
    \int Dz \frac{e^{-K} \int Dy 
    \tilde{\Xi}(h, E)  h^2}
    {1+e^{-K} \int Dy \tilde{\Xi}(h, E)}, \\
    \chi &= \frac{\rho}{E}, \\
    q_0 &= \frac{1}{E^2}\int Dz \left (
    \frac{e^{-K}\int Dy \tilde{\Xi}(h, E) h}
    {1+e^{-K} \int Dy \tilde{\Xi}(h, E)} \right )^2
    \label{eq:RS_SP_q0_3}, \\
    \rho &= \int Dz \frac{e^{-K}\int Dy \tilde{\Xi}(h, E)}{1+e^{-K}\int Dy \tilde{\Xi}(h, E)}.
\end{align} 
Using the solution of these equations, 
Eq.~(\ref{eq:typical_entropy}) is expressed as
\begin{align*}
    \Sigma =&\alpha \int Dz \ln
    \left [\int Dy\tilde{H}(v, \chi) \right ]
    + \int Dz \ln\left [
    1+e^{-K}\int Dy \tilde{\Xi}(h, E) 
    \right ]\cr
    &+   K\rho +\frac{1}{2}(E-F_1)\rho -\frac{F_1\chi}{2}+\frac{F_0q_0}{2}.
\end{align*}

\begin{figure}
\centering
\includegraphics[keepaspectratio,width=14cm]{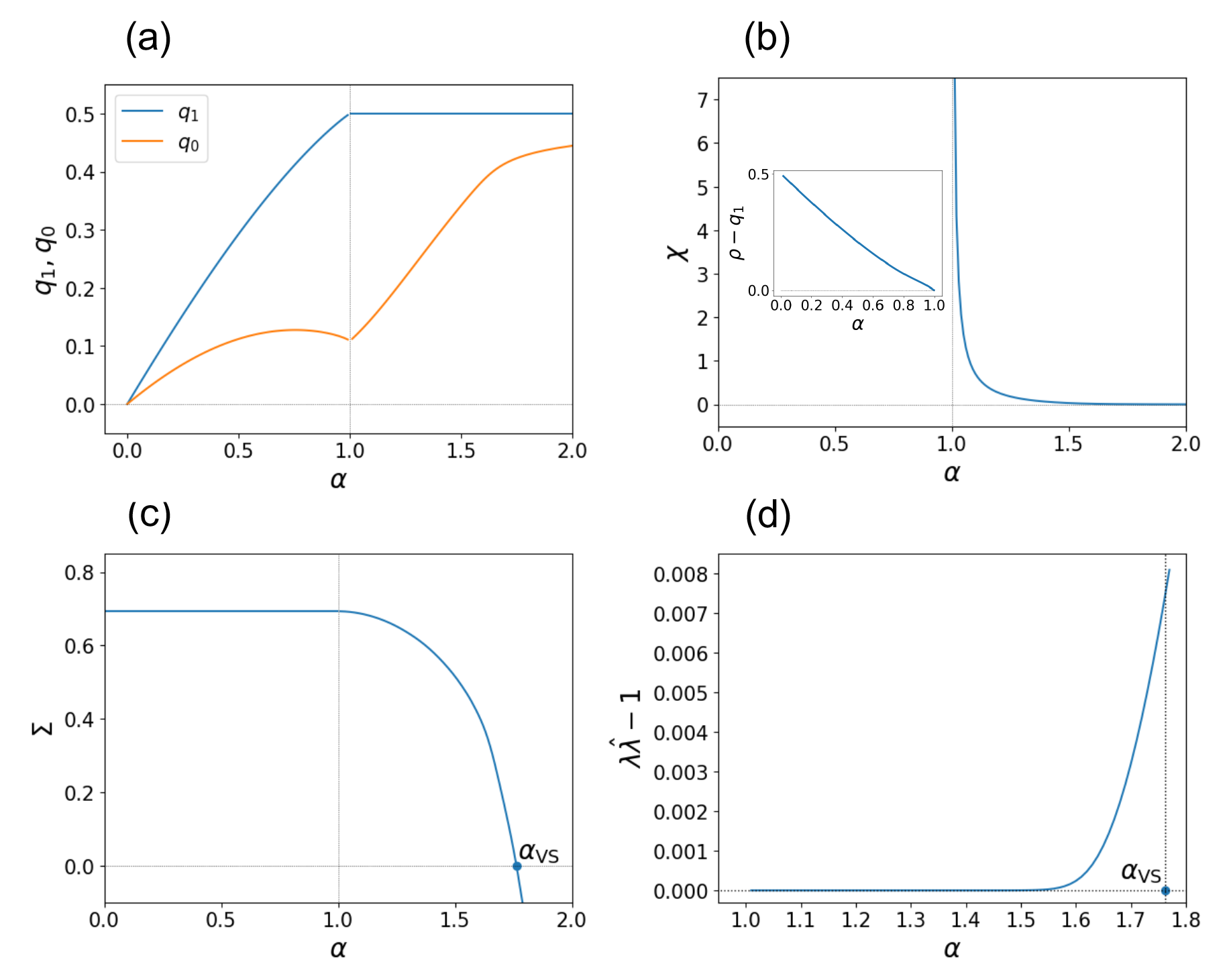}
\caption{\label{fig:graphs}
Profiles of the RS solutions for $\rho = 0.5$ are shown in  (a) $q_{1}$ and $q_{0}$,  
(b) $\chi$,  
(c) $\Sigma$, 
and 
(d) the AT stability condition (\ref{eq:AT})
as functions of $\alpha$. 
}
\end{figure}

We solved the equations numerically.
As a representative case, we plot relevant quantities in Fig. \ref{fig:graphs} together with those
for $\alpha < \alpha_{\rm CG}$
for $\rho = 0.5$.  
The vector $\bm{c}$, which specifies a set of the selected variables, serves as a label for solution clusters.
The quantity $q_{1}$ represents the typical similarity between parameter vectors $\bm{w}$ within the same cluster, whereas $q_{0}$ characterizes their typical similarity across different clusters.
Figure~\ref{fig:graphs}~(a) shows the dependence of $q_{1}$ and $q_{0}$ on $\alpha$.
As $\alpha$ approaches the Cover--Gardner capacity $\alpha_{\rm CG}$ from below, $q_{1}$ increases monotonically and converges to $\rho$ at $\alpha=\alpha_{\rm CG}$.
This indicates that, as the number of random patterns $P$ increases, the volume of the  feasible region within a typical cluster shrinks and eventually vanishes at $\alpha=\alpha_{\rm CG}$.
In contrast, $q_0$ exhibits a nontrivial behavior: it increases up to a maximum and then decreases for $\alpha < \alpha_{\rm CG}$, while for $\alpha > \alpha_{\rm CG}$ it increases monotonically.

Figure~\ref{fig:graphs}~(b) plots
\(
\chi = \lim_{m\to 0} m^{-1}(\rho - q_{0}).
\)
Since $\rho - q_{1}$ remains finite for $\alpha < \alpha_{\rm CG}$ (inset), $\chi$ takes a finite value only for $\alpha > \alpha_{\rm CG}$.
Figure~\ref{fig:graphs}~(c) shows the entropy density $\Sigma$ of clusters whose feasible region 
does not vanish.
At $\alpha=\alpha_{\rm CG}$, clusters 
in which no $\bm{w}$ is compatible with $\xi^P$
begin to emerge; as $\alpha$ increases further, $\Sigma$ decreases from the binary entropy
\(
-(1-\rho)\ln(1-\rho) - \rho \ln \rho,
\)
associated with the variable selection ratio 
$\rho$, and eventually reaches zero at some value $\alpha_{\rm VS}$.
This means that no clusters contain 
$\bm{w}$ compatible with $\xi^P$,  and this value $\alpha_{\rm VS}$ determines the capacity under variable selection.

The solution would provide an exact estimate of $\alpha_{\rm VS}$ if the replica symmetric (RS) assumption were valid. Unfortunately, this is not the case. As is well known, the RS solution must satisfy the de Almeida–Thouless (AT) stability condition, which requires that perturbations breaking replica symmetry do not grow~\citep{Almeida1978,Mezard1987}:
\begin{align}
\lambda\hat{\lambda} - 1 < 0 .
\label{eq:AT}
\end{align}
Here,
\begin{align*}
\lambda
&= \alpha \int Dz
\frac{\int Dy \tilde{H}(v,\chi)
\left(\frac{\partial^2}{\partial v^2}\ln \tilde{H}(v,\chi)\right)^2}
{\int Dy \tilde{H}(v,\chi)}
= \frac{\alpha}{\chi^2}
\int Dz
\frac{\int Dy \tilde{H}(v,\chi)\Theta(v)}
{\int Dy \tilde{H}(v,\chi)}, \\
\hat{\lambda}
&= \int Dz
\frac{ e^{-K}\int Dy \tilde{\Xi}(h, E)
\left(\frac{\partial^2}{\partial h^2}\ln \tilde{\Xi}(h,E)\right)^2}
{1 + e^{-K}\int Dy \tilde{\Xi}(h, E)}
= \frac{1}{E^2}
\int Dz
\frac{e^{-K}\int Dy \tilde{\Xi}(h, E)}
{1 + e^{-K}\int Dy \tilde{\Xi}(h,E)} .
\end{align*}
The derivation of this condition is provided in Appendix~\ref{app:AT}.
Figure~\ref{fig:graphs}~(d) shows that the AT condition \eqref{eq:AT} is violated for $\alpha > \alpha_{\rm CG}$, indicating that replica symmetry breaking (RSB) must be taken into account in this region, whereas the RS solution remains locally stable for $\alpha < \alpha_{\rm CG}$. This behavior is observed not only for $\rho = 0.5$ but also for other values of $\rho$. Therefore, the present results should be regarded as approximate solutions under the RS assumption.

Nonetheless, in many related problems, RS solutions—while quantitatively imperfect—are known to capture the qualitative behavior correctly~\citep{Amit1985,Fu1986,Monasson1995}. In this spirit, although quantitative discrepancies may remain, we expect that our analysis still correctly describes the qualitative scenario.

\section{Experimental verification}
To verify the above analytical results, we conducted numerical experiments.
Unfortunately, performing optimal variable selection is computationally intractable.
Therefore, our numerical study based on a heuristic algorithm called Iterative Hard Thresholding (BIHT) \citep{BIHTjacques2013} was aimed solely at demonstrating that, even for $\alpha > \alpha_{\rm CG}$, linear separability of random patterns by the perceptron becomes possible when variable selection is performed.

Given an initial estimate $\bm{x}^{0}=\bm{0}$ and the 1-bit measurements $\bar{\bm{y}}$, BIHT updates at iteration $l$ according to
\begin{eqnarray}
  \bm{a}_{l+1} & = & \bm{w}_{l}
  + \frac{\tau}{2} X^{T}\!\left( \bar{\bm{y}} - \mathrm{sign}\!\left(X \bm{w}_{l}\right) \right), 
  \label{BIHT1} \\[3pt]
  \bm{w}_{l+1} & = & \eta_{K}(\bm{a}_{l+1}),
  \label{BIHT2}
\end{eqnarray}
where $\tau$ is a step-size parameter controlling the gradient descent update, and 
$\eta_{K}(\bm{v})$ denotes 
the $K$-largest in magnitude components of $\bm{v}$
obtained by hard thresholding.
Once the algorithm terminates (either upon achieving consistency or reaching the maximum number of iterations), the final estimate is normalized to lie on the unit sphere.

The key to understanding BIHT lies in its underlying objective function.
As shown in \citep{BIHTjacques2013}, the update in Eq.~\eqref{BIHT1} corresponds to the negative subgradient of the convex objective
\[
{\cal J}(\bm{x})
= \left\| \bigl[\, \bar{\bm{y}} \odot (\Phi \bm{x}) \,\bigr]_{-} \right\|_{1}.
\]
Here, $[\cdot]_{-}$ denotes the negative part operator, defined component-wise as
\[
\left([\bm{u}]_{-}\right)_{i} = [u_{i}]_{-}, \qquad
[u_{i}]_{-} =
\begin{cases}
u_{i}, & u_{i} < 0, \\
0,     & \text{otherwise},
\end{cases}
\]
and $\bm{u}\odot\bm{v}$ denotes the Hadamard (element-wise) product,
\[
(\bm{u}\odot\bm{v})_{i} = u_{i} v_{i},
\]
for vectors $\bm{u}$ and $\bm{v}$.

\begin{algorithm}[t]
\caption{Greedy Binary Iterative Hard Thresholding (greedy-BIHT) algorithm}
\label{algGreedyBIHT}
\begin{algorithmic}[1]

\State \textbf{Given:} data set $\bm{y}$ and matrix $X$

\State \textbf{Initialization:}
\State Initialize $\bm{w}$ with i.i.d.\ Gaussian entries and rescale by $\sqrt{N}$
\State Set the number of nonzero coefficients $K \gets 1$
\State Set counter $l \gets 0$

\While{\text{err} $> \epsilon$ \textbf{and} $l \leq L$}
    \State $l \gets l + 1$
    \State $\bm{a}_{l} \gets \bm{w}_{l-1} 
    + \frac{\tau}{2} X^{T}\!\left( \bm{y} - \mathrm{sign}\!\left( X \bm{w}_{l-1} \right)\right)$
    \State $\bm{w}_{l} \gets \eta_{K}(\bm{a}_{l})$
    \State $\text{err} \gets N^{-1}\|\bm{w}_{l} - \bm{w}_{l-1}\|_{2}$
\EndWhile

\State $\Delta \gets \bm{y} \odot (X \bm{w})$
\State Find the support of nonzero entries: $I_{f} \gets f_{\setminus 0}(\bm{w})$

\While{$\mathrm{sum}(\Delta < 0) > 0$ \textbf{and} $K < N$}
    \State $K \gets K + 1$
    \While{\text{err} $> \epsilon$ \textbf{and} $l \leq L$}
        \State $l \gets l + 1$
        \State $\bm{a}_{l} \gets \bm{w}_{l-1} 
        + \frac{\tau}{2} X^{T}\!\left( \bm{y} - \mathrm{sign}\!\left( X \bm{w}_{l-1} \right)\right)$
        \State $\bm{w}_{l} \gets \eta_{K}(\bm{a}_{l} \mid I_{f})$
        \State $\text{err} \gets N^{-1}\|\bm{w}_{l} - \bm{w}_{l-1}\|_{2}$
    \EndWhile
    \State $I_{f} \gets f_{\setminus 0}(\bm{w})$
    \State $\Delta \gets \bm{y} \odot (X \bm{w})$
\EndWhile

\State \textbf{return} $\bm{w}$

\end{algorithmic}
\end{algorithm}

In our experiments, we developed a greedy search procedure based on the BIHT algorithm, which we refer to as the \emph{greedy-BIHT} algorithm. The pseudocode is summarized in Algorithm~\ref{algGreedyBIHT}.
The function $\mathrm{sum}(\Delta < 0)$ counts the number of entries in the vector $\Delta$ that are negative, corresponding to the mismatched 
data. 
The function $f_{\setminus 0}(\bm{v})$ returns an indicator vector specifying the nonzero components of $\bm{v}$. 
The condition ``\textbf{while} $\mathrm{sum}(\Delta < 0) > 0$ \textbf{and} $K < N$'' therefore means that, as long as there exist mismatches between the predicted signs and the output vector $\bm{y}$, and the number of nonzero components in the weight vector $\bm{w}$ can still be increased, the algorithm continues to update within the while-loop.

The operator $\eta_{K}(\bm{v}\mid I_{f})$ fixes the positions indicated by $I_{f}$ and selects the remaining nonzero entries by sorting the absolute values of the unfixed components, keeping only the $K$ largest in magnitude and setting the rest to zero. 
In the greedy-BIHT algorithm, since $K$ is increased one by one, the update 
\[
\bm{w}_{l} \gets \eta_{K}(\bm{a}_{l} \mid I_{f})
\]
keeps the nonzero entries already present in $\bm{w}_{l-1}$ and adds one additional nonzero position in $\bm{a}_{l}$ with the largest absolute value, while setting all other components to zero.

\begin{figure}
\centering
\includegraphics[keepaspectratio,width=12cm]{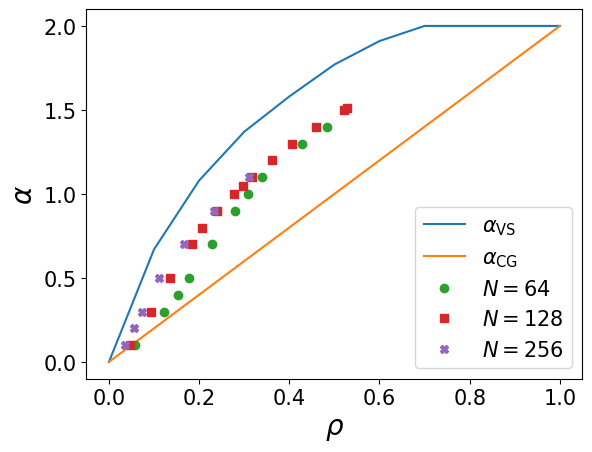}
\caption{
Perceptron capacity as a function of the variable selection ratio $\rho$. 
The solid blue line represents the capacity $\alpha_{\rm VS}$ predicted by the replica symmetric (RS) analysis under optimal variable selection, 
while the solid orange line shows the classical Cover--Gardner capacity $\alpha_{\rm CG} = 2\rho$. 
The green, red, and purple markers correspond to the averaged results of the greedy-BIHT experiments for system sizes $N = 64, 128, 256$, respectively.
}
\label{fig:capacityphasediagram_alpha_rho_greedy}
\end{figure}

In the experiments, we set the BIHT gradient parameter to $\tau = 0.002/P$, the termination threshold for the weight-update error to $\epsilon = 10^{-8}$, and the maximum number of iterations to $L=1000$. 
We performed simulations for finite-size systems with dimensions $N = 64, 128, 256$. 
For each dimension, we conducted $T = 1000$ independent trials. 
By averaging over these trials, we obtained the plots shown in Fig.~\ref{fig:capacityphasediagram_alpha_rho_greedy}.
As $\alpha$ increases, searching for a solution becomes more difficult.
Consequently, with our available computational resources, we were not able to obtain solutions for $\alpha \gtrsim 1.5$.
Although there is a discrepancy between the experimental results and the capacity predicted by the RS solution, our findings clearly demonstrate that variable selection enables the perceptron to separate random patterns even beyond the Cover--Gardner capacity $\alpha_{\rm CG} = 2\rho$.

\section{Conclusion and discussion}




In this work, we investigated the storage capacity of a perceptron under
variable selection, where only a subset $M=\rho N$ of variables among $N$
candidates can be used to classify $P=\alpha N$ patterns. This allows us
to quantitatively address a central question in binary classification:
when variable selection is performed, how many patterns must be
correctly classified before we can interpret the learned classifier as
capturing genuine structure rather than merely fitting chance correlations?
By establishing a theoretical boundary separating classification success
from failure for random datasets, our analysis provides a principled criterion for determining
when the set of selected variables represents essential information in high-dimensional data.

We also developed a simple greedy-BIHT algorithm in the same setting.
Although it does not achieve the theoretical limit, its capability to
correctly classify more patterns than the classical Cover--Gardner
capacity under variable selection supports the validity and practical
relevance of the theoretical findings, and illustrates that algorithmic
learning remains feasible even under stringent sparsity constraints.

Besides, our theoretical results clarify how the restriction on the number
of couplings influences the maximal number of retrievable memory patterns in
associative memory models, indicating that more patterns can be stored per
coupling than the result known before in networks of asymmetric couplings.
This result suggests that appropriate selection of the connectivity structure within the network can lead to robustness against performance degradation when the number of synaptic connections is limited.

Overall, our study provides a unified theoretical perspective on how
variable selection governs the ability to distinguish structure from noise
in binary classification while simultaneously constraining memory storage
in networks with limited connectivity. We hope that these findings
stimulate further development of resource-efficient learning algorithms
and deepen the interplay between statistical mechanics and modern machine
intelligence.

\section*{Acknowledgements}
An early version of this work was presented at the Physical Society of Japan (JPS) 2015 Fall Annual Meeting, held in Japan, and at the Statistical Physics and Neural Computation conference, held on October 4-6, 2019, at Sun Yat-sen University, China.
Y. X. acknowledges support from the Finnish Center for Artificial Intelligence (FCAI) and the computational resources provided by CSC – IT Center for Science, Finland, via the Puhti supercomputing infrastructure.
M. O. received financial support from the programs for Bridging the gap between R\&D and the IDeal society (society 5.0), Generating Economic and social value (BRIDGE), and the Cross-ministerial Strategic Innovation Promotion Program (SIP) from the Cabinet Office.
Y. K. acknowledges support from 
MEXT/JSPS KAKENHI Grant No. 22H05117.

\appendix

\section{Derivation of Eq.~(\ref{eq:RS_generating_function})}
\label{app:RS_computation}
When the replica-symmetric assumption (\ref{eq:RS_assumption}) holds,  
the Gaussian random variables satisfying Eq.~(\ref{eq:covariances}) can be written as  
\begin{align*}
\numberthis \label{eq:Gaussian_decomposition}
u^{a\sigma}
= \sqrt{\rho - q_{1}}\, x^{a\sigma}
  + \sqrt{q_{1} - q_{0}}\, y^{a}
  + \sqrt{q_{0}}\, z,
\qquad
(a = 1,\ldots,n;\ \sigma = 1,\ldots,m),
\end{align*}
using standard Gaussian variables $x^{a\sigma}$, $y^{a}$, and $z$, which are mutually independent.
This decomposition yields
\begin{align*}
& \int 
      \frac{
        \prod_{a=1}^{n}\prod_{\sigma=1}^{m} 
        du^{a\sigma}\;
        \exp\!\left(
          -\frac{1}{2}\bm{u}^{\top}\mathcal{Q}^{-1}\bm{u}
        \right)
      }{
        (2\pi)^{nm/2} \bigl(\det \mathcal{Q}\bigr)^{1/2}
      }
      \prod_{a=1}^{n}\prod_{\sigma=1}^{m}\Theta(u^{a\sigma})
      \nonumber\\
&\quad =
        \int Dz 
        \left(
          \prod_{a=1}^{n}
          \int Dy^{a}
          \prod_{\sigma=1}^{m}
            \int Dx^{a\sigma}\,
              \Theta\!\left(
                  \sqrt{\rho - q_{1}}\,x^{a\sigma}
                + \sqrt{q_{1} - q_{0}}\,y^{a}
                + \sqrt{q_{0}}\,z
              \right)
        \right)
      \nonumber\\
&\quad =
        \int Dz 
        \left(
          \int Dy\, 
            H^{m}\!\left(
                -\frac{\sqrt{q_{1}-q_{0}}\,y 
                       + \sqrt{q_{0}}\,z}{\sqrt{\rho - q_{1}}}
            \right)
        \right)^{n}
      \nonumber\\
&\quad =
        \int Dz 
        \left(
          \int Dy\, 
            H^{m}\!\left(
                \frac{\sqrt{q_{1}-q_{0}}\,y 
                      + \sqrt{q_{0}}\,z}{\sqrt{\rho - q_{1}}}
            \right)
        \right)^{n}.
\end{align*}
A standard Gaussian identity,
\begin{align*}
\exp\!\left(\frac{b^{2}}{2}\right)
= \int Dz\, e^{bz},
\end{align*}
leads to
\begin{align*}
& \frac{1}{(2\pi)^{nm/2}}
  \sum_{c^{1},\ldots,c^{n}}
  \int 
    \prod_{a=1}^{n}\prod_{\sigma=1}^{m} 
      dw^{a\sigma}\,
  \exp\!\left(
      -\sum_{a=1}^{n}K_{a}c^{a}
      + \mathcal{L}
           (\{c^{a}\}, \{w^{a\sigma}\}, 
            \{\hat{q}_{ab;\sigma\tau}\})
   \right)
   \nonumber\\
&\quad =
  \frac{1}{(2\pi)^{nm/2}}
  \int Dz\,
  \left(
     \sum_{c^{a}}
       e^{-Kc^{a}}
       \int Dy^{a}
       \prod_{\sigma=1}^{m}
         \int dw^{a\sigma}
           \exp\!\left(
             \mathcal{L}^{\mathrm{RS}}
           \right)
  \right)^{n}
  \nonumber\\
&\quad =
  \int Dz
  \prod_{a=1}^{n}
    \left(
      \sum_{c^{a}}
      \frac{e^{-Kc^{a}}}{(\hat{Q}+\hat{q}_{1})^{mc^{a}/2}}
      \int Dy^{a}
      \exp\!\left(
        \frac{
          m c^{a}\bigl(\sqrt{\hat{q}_{1}-\hat{q}_{0}}\,y^{a}
                  +     \sqrt{\hat{q}_{0}}\,z
             \bigr)^{2}
        }{
          2(\hat{Q}+\hat{q}_{1})
        }
      \right)
    \right)
  \nonumber\\
&\quad =
  \int Dz\,
  \left(
    1 
    + \frac{e^{-K}}{(\hat{Q}+\hat{q}_{1})^{m/2}}
        \int Dy\,
        \exp\!\left(
          \frac{
            m\bigl(\sqrt{\hat{q}_{1}-\hat{q}_{0}}\,y
                 + \sqrt{\hat{q}_{0}}\,z
            \bigr)^{2}
          }{
            2(\hat{Q}+\hat{q}_{1})
          }
        \right)
  \right)^{n}, 
\end{align*}
where 
\begin{align*}
{\mathcal L}^{\mathrm{RS}}
&=
  \sum_{a=1}^{n}
  \frac{1 - c^{a}}{2}
   \sum_{\sigma=1}^{m} (w^{a\sigma})^{2}
  \nonumber\\
&\quad
 + \sum_{a=1}^{n}\sum_{\sigma=1}^{m}
    \left(
       -\frac{\hat{Q}+\hat{q}_{1}}{2}
        (c^{a} w^{a\sigma})^{2}
       + \bigl(\sqrt{\hat{q}_{1}-\hat{q}_{0}}\,y^{a}
              + \sqrt{\hat{q}_{0}}\,z
         \bigr)
         (c^{a} w^{a\sigma})
    \right).
\end{align*}
Counting the number of combinations yields
\begin{align*}
& \rho \sum_{a=1}^{n} K_{a}
  + \rho \sum_{a=1}^{n}\sum_{\sigma=1}^{m}
      \frac{\hat{q}_{aa;\sigma\sigma}}{2}
  - \sum_{\substack{
        a\le b,\; \sigma\le \tau \\
        a\ne b\ \vee\ \sigma\ne\tau
      }}
     \hat{q}_{ab;\sigma\tau} q_{ab;\sigma\tau}
     \nonumber\\
&\qquad
  =
  nK\rho
  + \frac{nm}{2}\hat{Q}\rho
  - \frac{nm(m-1)}{2}\bigl(\hat{q}_{1}q_{1}-\hat{q}_{0}q_{0}\bigr)
  - \frac{nm(nm-1)}{2}\hat{q}_{0}q_{0}.
\end{align*}
Substituting all expressions above into  
Eq.~(\ref{eq:logV_integers_n_m}) provides  
Eq.~(\ref{eq:RS_generating_function}).

\section{Derivation of Eq.~(\ref{eq:AT})}
\label{app:AT}
The one-step replica symmetry breaking (1RSB) solution is constructed by 
partitioning the $m$ replica indices $\{1,\ldots,m\}$ into $m/k$ groups of equal size $k$, 
and assuming the following structure:
\begin{align*}
    q_{ab;\sigma\tau} =\left \{
    \begin{array}{ll}
    \rho, & a=b, \sigma = \tau, \\
    q_2, & a=b, \mbox{$\sigma$ and $\rho$ are in a same group}, \\
    q_1, & a=b, \mbox{$\sigma$ and $\rho$ are not in a same group}, \\
    q_0, & a\ne b. 
    \end{array}
    \right .
\end{align*}
A similar ansatz is also assumed for $\hat{q}_{ab;\sigma\tau}$. Under this assumption, the saddlepoint condi-
tion becomes
\begin{align*}
    \hat{q}_2 &= \alpha\int Dz \frac{\int Dy \left (\int Dx H^k \right )^{m/k}
    \frac{\int Dx H^k  
    \left (\frac{H^\prime}{H} \right )^2}{\int Dx H^k} }
    {\int Dy \left (\int Dx H^k \right )^{m/k}}, \\
    \hat{q}_1 &= \alpha \int Dz \frac{\int Dy \left (\int Dx H^k \right )^{m/k}
    \left (\frac{\int Dx H^k
\frac{H^\prime}{H}}{\int DxH^k} \right )^2 }{\int Dy \left (\int Dx H^k \right )^{m/k}}, \\
\hat{q}_0 &= \alpha \int Dz \left (\frac{\int Dy \left (\int Dx H^k \right )^{m/k}
    \frac{\int Dx H^k
\frac{H^\prime}{H}}{\int DxH^k} }{\int Dy \left (\int Dx H^k \right )^{m/k}} \right )^2, \\
\rho &=\frac{\rho}{\hat{Q}+\hat{q}_2} + q_2 \\
q_2 &= \int Dz \frac{e^{-K}\int Dy \left (\int Dx \Xi^k \right )^{m/k} 
\frac{\int Dx \Xi^k \omega^2 }{\int Dx \Xi^k}}{
1+ e^{-K}\int Dy \left (\int Dx \Xi^k \right )^{m/k} }, \\
q_1 &= \int Dz \frac{e^{-K}\int Dy \left (\int Dx \Xi^k \right )^{m/k} 
\left (\frac{\int Dx \Xi^k \omega }{\int Dx \Xi^k}\right )^2}{
1+ e^{-K}\int Dy \left (\int Dx \Xi^k \right )^{m/k} }, \\
q_0 &= \int Dz \left (\frac{e^{-K}\int Dy \left (\int Dx \Xi^k \right )^{m/k} 
\frac{\int Dx \Xi^k \omega }{\int Dx \Xi^k}}{
1+ e^{-K}\int Dy \left (\int Dx \Xi^k \right )^{m/k} }\right )^2, \\
\rho &= \int Dz \frac{e^{-K}\int Dy \left (\int Dx \Xi^k \right )^{m/k} }{
1+ e^{-K}\int Dy \left (\int Dx \Xi^k \right )^{m/k} }, 
\end{align*}
where $H= H\left (\frac{v}{\sqrt{\rho-q_2}} \right )$, $v=\sqrt{q_2-q_1}x + \sqrt{q_1-q_0} y + \sqrt{q_0}z$, 
$\Xi = (\hat{Q}+\hat{q}_2)^{-1/2} \exp \left (\frac{h^2}{2(\hat{Q}+\hat{q}_2)} \right )$, 
$\omega  = h/(\hat{Q}+\hat{q}_2)$, and $h=\sqrt{\hat{q}_2 -\hat{q}_1}x + \sqrt{\hat{q}_1-\hat{q}_0}y + \sqrt{\hat{q}_0} z$. 

The RS solution is recovered as a special case of the 1RSB solution by imposing $q_2=q_1$
and $\hat{q}_2 = \hat{q}_1$. To investigate its local stability, we introduce perturbations 
$\Delta = q_2-q_1$ and $\hat{\Delta} = \hat{q}_2 -\hat{q}_1$ and linearize the 
above equations around the RS solution. This yields
\begin{align*}
    \hat{\Delta} &\simeq \alpha \int Dz \frac{\int Dy \left (\int Dx H^k \right )^{m/k}
\frac{\int Dx H^k 
    \left ( \left (\frac{\partial^2}{\partial v^2} \ln H \right )\sqrt{\Delta} x\right )^2}
    {\int Dx H^k}}
    { \int Dy \left (\int Dx H^k \right )^{m/k}} \cr
    &= \alpha \int Dz \frac{\int Dy H^m\left (\frac{\partial^2}{\partial v^2} \ln H \right )}
    { \int Dy H^m } \Delta, \\
    \Delta &\simeq \int Dz \frac{e^{-K}\int Dy \left (\int Dx \Xi^k \right )^{m/k} 
\frac{\int Dx \Xi^k \left( \left (\frac{\partial^2}{\partial h^2}\ln \Xi \right )\sqrt{\hat{\Delta}} x\right )^2 }{\int Dx \Xi^k}}{
1+ e^{-K}\int Dz \int Dy \left (\int Dx \Xi^k \right )^{m/k} } \cr
&=  \int Dz \frac{e^{-K}\int Dy  \Xi^m  \left (\frac{\partial^2}{\partial h^2}\ln \Xi \right )^2 }{
1+ e^{-K}\int Dy \Xi^m } \hat{\Delta}, 
\end{align*}
where we used the fact that $v$ and $h$ do not depend on variable $x$ when $q_2=q_1$ and $\hat{q}_2 = \hat{q}_1$ 
hold and $\int Dx x^2 = 1$. The linearized equations offer the local stability condition of the RS solution as
\begin{align*}
    \alpha \int Dz \frac{\int Dy H^m\left (\frac{\partial^2}{\partial v^2} \ln H \right )}
    { \int Dy H^m } \times \int Dz \frac{e^{-K}\int Dy  \Xi^m  \left (\frac{\partial^2}{\partial h^2}\ln \Xi \right )^2 }{
1+ e^{-K}\int Dy \Xi^m } < 1.  
\end{align*}
Taking the limit of $m\to 0$, together with the appropriate rescaling, leads to Eq.~(\ref{eq:AT}).

\newpage


\end{document}